\documentclass[aps,prb,showpacs,twocolumn,superscriptaddress]{revtex4}

\usepackage{graphicx,bm,amssymb}

\begin{document}
\title{Phase behavior
and interfacial properties \\of nonadditive mixtures of Onsager
rods}
\author{Kostya Shundyak}
\affiliation {Institute for Theoretical Physics, Utrecht
University, Leuvenlaan 4, 3584 CE Utrecht, The Netherlands}
\author{Ren\'e van Roij}
\affiliation {Institute for Theoretical Physics, Utrecht
University, Leuvenlaan 4, 3584 CE Utrecht, The Netherlands}
\author{Paul van der Schoot}
\affiliation {Eindhoven Polymer Laboratories, Eindhoven University
of Technology,\\ P.O. Box 513, 5600 MB Eindhoven, The Netherlands}
\date{December 6, 2004}

\begin{abstract}
Within a second virial theory, we study bulk phase diagrams as
well as the free planar isotropic-nematic interface of binary
mixtures of nonadditive thin and thick hard rods. For species of
the same type the excluded volume is determined only by the
dimensions of the particles, whereas for dissimilar ones it is
taken to be larger or smaller than that, giving rise to a
nonadditivity that can be positive or negative. We argue that such
a nonadditivity can result from modelling of soft interactions as
effective hard-core interactions. The nonadditivity enhances or
reduces the fractionation at isotropic-nematic ($IN$) coexistence
and may induce or suppress a demixing of the high-density nematic
phase into two nematic phases of different composition ($N_1$ and
$N_2$), depending on whether the nonadditivity is positive or
negative. The interfacial tension between co-existing isotropic
and nematic phases show an increase with increasing fractionation
at the $IN$ interface, and complete wetting of the $IN_2$
interface by the $N_1$ phase upon approach of the triple point
coexistence. In all explored cases bulk and interfacial properties
of the nonadditive mixtures exhibit a striking and quite
unexpected similarity with the properties of additive mixtures of
different diameter ratio.
\end{abstract}

\pacs{61.30.Cz, 64.70.Md, 05.70.Np} \maketitle

\section{Introduction}
In his paper about the isotropic-nematic ($IN$) transition in
solutions of monodisperse, rod-like particles that interact
through a hard, steric repulsion, Onsager briefly discussed a
possible extension of his results to polydisperse systems
\cite{O}. Since then, a tremendous amount of work has been devoted
to the study of the influence of polydispersity on the phase
behavior of such hard-rod fluids, both for the case where this
polydispersity is of the quenched type \cite{Vroege,Sollich,Chen, Yuri} and
for where it is of the annealed type \cite{Taylor,Kramer,Schoot,Baulin}.
Focusing on the former, even the simplest (binary) mixtures
consisting of long hard rods that differ only in length or
diameter exhibit quite nontrivial phase diagrams. In addition to
the pure isotropic and nematic phases of various composition and
regions of their coexistence, the high-density nematic phase can
demix (and possibly remix) into two nematic phases of different
composition (denoted $N_1$ and $N_2$). The reason for the
existence of an $IN$ transition in binary mixtures is the same as
that in a monodisperse hard-rod fluids, being a competition
between orientation entropy and entropy of packing \cite{O,VL}. In
contrast, the nematic-nematic demixing does not involve changes in
excluded volume (i.e. packing entropy), but rather a competition
between orientation entropy and entropy of mixing \cite{VLJPC93}.
Another interesting feature is that for sufficiently large size
disparity, the two distinct nematic phases do not remix even at
arbitrary high pressure \cite{RMEL96}.

Unfortunately, it is quite difficult to compare these theoretical
findings with results obtained from actual experiments. Although
rod-like particles can be synthesized chemically in various ways
\cite{VL}, typically their size distribution is mono- or
bi-disperse only to a first approximation. By contrast,
suspensions of rod-like viruses such as tobacco mosaic virus, M13,
pf1 and fd are characterized by a high degree of monodispersity,
and are therefore attractive model systems, despite the
complicating factors associated with their fixed physical
dimensions, their charged nature and the fact that they are not
actually infinitely rigid but exhibit some degree of bending
flexibility.

Recently, however, experimental procedures have been developed
that allow one to modify the length and the diameter of these
viruses \cite{DFPTRSLA01}, which opens the possibility to form
binary mixtures of a well-defined bidispersity. In particular, one
of the methods is based on altering the effective diameter of the
fd-virus by coating it with the polymer polyethylene glycol (PEG).
Studies of such binary mixtures of thin and thick rods have
revealed coexistence regions of the isotropic and different
nematic phases ($IN_2$ and $IN_1$), as well as a nematic-nematic
coexistence region ($N_1N_2$) and an $IN_1N_2$ triple point
\cite{PVGJFcondmat}. Although some of the gross features of this
experimentally determined phase diagram are in agreement with
theoretical predictions based on an extension of Onsager's
second-virial theory to binary mixtures of hard rods
\cite{RMDP98,SRPRE03}, some of the experimental and theoretical
findings turn out to be in sharp contrast with each other.

According to the theory, mixtures of thin rods (with a diameter
$D_1$) and thick ones (diameter $D_2$) of equal length $L$ should
exhibit a spindle-like $IN$ coexistence without any
nematic-nematic demixing for diameter ratios $d=D_2/D_1<3.8$.
Experiments, however, point at a broad $N_1N_2$ coexistence for a
diameter ratio as small as $d\sim 2.0$ \cite{PVGJFcondmat}.
Furthermore, in the interval $3.8<d<4.29$ the single nematic phase
demixes according to the theory into two nematic phases $N_1$ and
$N_2$ of different composition, whilst remixing takes place at
sufficiently high total density (or osmotic pressure), that is,
above an {\em upper} critical (or consolute) point. Experiments,
however, reveal a {\em lower} consolute point that closes the
$N_1-N_2$ coexistence \cite{PVGJFcondmat}, i.e., the $N_1-N_2$
demixing becomes more pronounced with increasing osmotic pressure.

Possible explanations for these differences may well be found in
the idealisations incurred when modelling the virus particles as
infinitely elongated, infinitely rigid rods that interact with
each other only through additive hard-core potentials. Indeed, the
virus particles are semi-flexible and charged, as already alluded
to. In addition, the grafted polymer coating is soft and hence
compressible, and the length-to-diameter ratio of the rods is at
best, say, 50. It is important to recall that the second virial
theory is believed to be exact only in the limit of infinite
aspect ratios of the rods \cite{VL}. The impact of a finite
length-to-diameter ratio was recently considered within an
extension of the so-called Parsons-Lee theory to mixtures of hard
rods \cite{VGJMP03}. This theory does reproduce a lower consolute
point for mixtures of thin and thick rods albeit only if their
length (presumed equal) is extremely small. A lower consolute
point has also been predicted for binary mixtures of semi-flexible
hard thin rods of unequal thickness, at least if their persistence
lengths and their widths do not differ by more than roughly the
square root of either persistence length over their contour length
\cite{semenov-subbotin,rvr}. However, the predicted
isotropic-isotropic demixing is not found in the experiments
involving the mixtures of naked and coated fd virus particles
\cite{PVGJFcondmat}. The theory \cite{semenov-subbotin,rvr} does
anyway not strictly apply to this experimental system, because of
the tacit assumption that the length of the rods greatly exceeds
their persistence length.

In an attempt to shed light on the issue, we focus on the effects
that any nonadditivity of the interactions between the two kinds
of rod might have on their phase behavior. Such a nonadditivity
emerges naturally if one replaces actual soft rod-rod repulsions
by effective hard-core repulsions, characterized by effective
hard-core diameters that in effect are distances of closest
approach. As is well known, the screened-Coulomb interactions
between charged virus particles in an electrolyte solution can be
reasonably approximated by an effective steric interaction with a
hard-core diameter that is the sum of the bare, "physical"
diameter of the rod and an electrostatic contribution proportional
to the Debye screening length of the suspending medium
\cite{SLOM86}. For the interaction between a pair of
polymer-coated virus particles one would have an effective
diameter of the order of the radius of gyration of the tethered
chains \cite{Grelet}, at least if the Debye length is much smaller
than that.

It is not at all obvious that the effective interaction length
between a bare and a polymer-coated charged rod should be the
linear average of the interaction lengths of the two separate
species, in other words, one would from the outset expect the
interaction within such an effective description to be
non-additive rather than additive. Indeed, as we shall see below
in section \ref{nonaddsect}, even highly simplified model
potentials produce non-additive effective hard-core interactions
in mixtures of rods. The level of nonadditivity may be expressed
in a parameter $\alpha$ defined such that the effective hard-core
diameter of an unlike pair of rods can written as
$\frac{1}{2}(D_1+D_2)(1+\alpha)$, where $D_{\sigma}$ is the
effective hard-core diameter of the interaction between two like
rods of species $\sigma=1,2$. For an additive mixute, $\alpha =0$.
In this paper, we make plausible by explicitly considering the
steric interactions between the various types of rod that, even
within a simplified model, $\alpha$ may attain values that can be
positive or negative up to, say, ten per cent. Additional sources
of nonadditivity may be found, say, in electric polarisation
effects of the charges on the polymer coating, but these will not
be considered here.

Although the microscopic origin of nonadditivity is ultimately
based on the less ($\alpha >0$) or more ($\alpha <0$) efficient
packing of the mixture compared to the pure species, we do not
attempt to calculate $\alpha$ from a realistic microscopic theory.
Having ascertained that $\alpha$ need indeed not be zero, we treat
it as a phenomenological parameter in a generalised Onsager
theory, and investigate its consequences for the phase behavior of
the mixture, and for the interfacial properties of co-existing
isotropic and nematic phases. As we shall see, both the predicted
phase diagrams and interfacial properties of the isotropic-nematic
interface are very sensitive to values of $|\alpha|$ as small as a
few per cent. Of course, this does not imply that all effects of
electrostatic interactions, flexibility, etc., are accurately or
even properly modelled. In fact, we find that non-additivity
cannot explain the existence of the lower consolute point found by
Fraden and co-workers \cite{PVGJFcondmat}.

The remainder of this paper is organized as follows. In Sec.
\ref{nonaddsect} we introduce a simple model for polymer-coated
rods, and provide an estimate for typical values of the
nonadditivity parameter $\alpha$. In Sec. \ref{nafunctsect} we
introduce the Onsager-type free energy functional, and derive from
that the basic Euler-Lagrange equations describing the
orientational and density distribution of the rods under
conditions of thermodynamic equilibrium. In Sec. \ref{nabulksect}
we solve these equations for bulk geometries, and analyze the
structure of a few typical bulk phase diagrams. In Sec.
\ref{nainterfsect} we briefly describe a method to solve the
Euler-Lagrange equation for interface geometries of binary
mixtures, and study $IN_1$, $N_1N_2$ and $IN_2$ interfaces, the
latter in particular in the vicinity of the bulk $IN_1N_2$ triple
point. A summary and discussion of the results are presented in
Sec. \ref{nasummarysect}.

\section{Nonadditivity of interactions}
\label{nonaddsect} The physical origin of nonadditivity can be
illustrated on the basis of a simple model for a mixture of bare
and PEG-coated fd viruses \cite{PVGJFcondmat}. The bare rods are
modelled as rigid hard rods of length $L$ and diameter $\Delta_1$
($L\gg\Delta_1$), and hence the interaction potential between two
bare rods, $\phi_{11}(r)$, is given by
$\beta\phi_{11}(r)=\infty,0$ for $r<\Delta_1$ and $r>\Delta_1$,
respectively. Here, $r$ denotes the shortest distance between the
main axes of the two rods.

The PEG-coated rods are identical to the bare ones, except that
they bear an additional soft layer extending to a distance
$\Delta_2/2$ from the axis of the rod, i.e., to a distance
$(\Delta_2-\Delta_1)/2$ from their hard-core surface. We do not
specify the relation between the dimensions of the tethered PEGs
and $\Delta_2$ in any detail, but one expects that
$(\Delta_2-\Delta_1)/2$ is of the order of the radius of gyration
of the grafted PEG (so we only consider $\Delta_2>\Delta_1$). We
expect that the soft, repulsive interaction that occurs when the
polymer coating of two rods overlap should be quite similar to
that of overlapping star polymers \cite{Lowen}. In order to keep
the model as simple as possible, we represent the interaction of
mean force resulting from the presence of a polymer coating by a
square-shoulder potential that is a function of $r$ alone, and
ignore any angle dependence that might arise in reality. This
angular dependence should be significant only for configurations
of rods inclined at small angles, which bear only a tiny
statistical weight in the limit of large aspect ratios. Note that
although our representation of the soft potential is isotropic,
the virials based on it are anisotropic because the interaction
volumes are a function of the relative orientations of the rods.

The interaction potential between a bare and a coated rod,
$\phi_{12}(r)$, should obviously be identical to the naked-rod
potential $\phi_{11}(r)$ if $r<\Delta_1$ and
$r>(\Delta_2+\Delta_1)/2$. Within our description, $\phi_{12}(r)$
takes on a value different from that, $\epsilon_1>0$, if
$\Delta_1<r<(\Delta_1+\Delta_2)/2$, i.e., when the hard-core of
the bare rod perturbes the soft outer layer of the coated rod. It
is to be seen as an average of the actual interaction potential
over its range. Our effective interaction potential between two
coated rods, $\phi_{22}(r)$, is more complicated and consists of
two shoulders in between the range of the hard-core repulsion
($r<\Delta_1$) and the noninteracting long-distance regime
($r>\Delta_2$). The first shoulder, for
$\Delta_1<r<(\Delta_1+\Delta_2)/2$, is such that
$\phi_{22}(r)=2\epsilon_1$, and represents the overlap of the hard
core of the first rod with the polymer layer of the second one,
and {\it vice versa} by symmetry. The second shoulder, for
$(\Delta_1+\Delta_2)/2<r<\Delta_2$, represents overlap of the two
polymer layers, and is such that $\phi_{22}(r)=\epsilon_2>0$.

\begin{figure}[!t]\centering
\includegraphics[width=8.5cm]{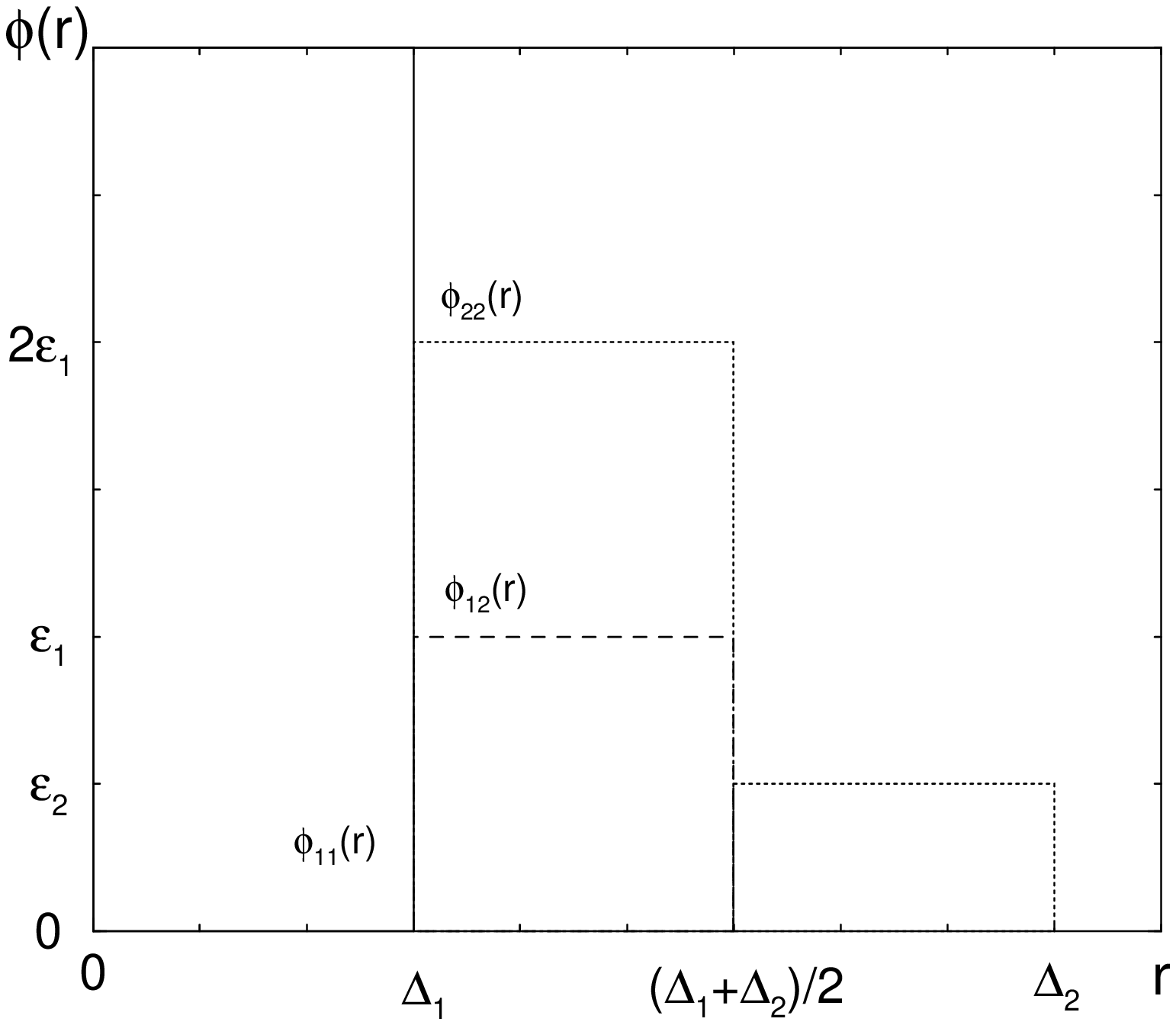}
\caption{\label{nonaddaalpha1}  K. Shundyak, R. van Roij and P. van der Schoot}
\end{figure}

The nature of the polymer chains is such that we expect their
entropy to be reduced more by a penetrating rigid rod than by
another polymer. Indeed, the cross virial of a rod and a flexible
chain is much larger than the geometric average of the rod-rod and
the chain-chain virials \cite{joanny}. For this reason we only
consider cases where $2\epsilon_1>\epsilon_2$. The pair potentials
$\phi_{\sigma\sigma'}(r)$ between rods of species $\sigma$ and
$\sigma'$ are illustrated graphically in Fig. \ref{nonaddaalpha1}.

\begin{figure}[!t]\centering
\includegraphics[width=8.5cm]{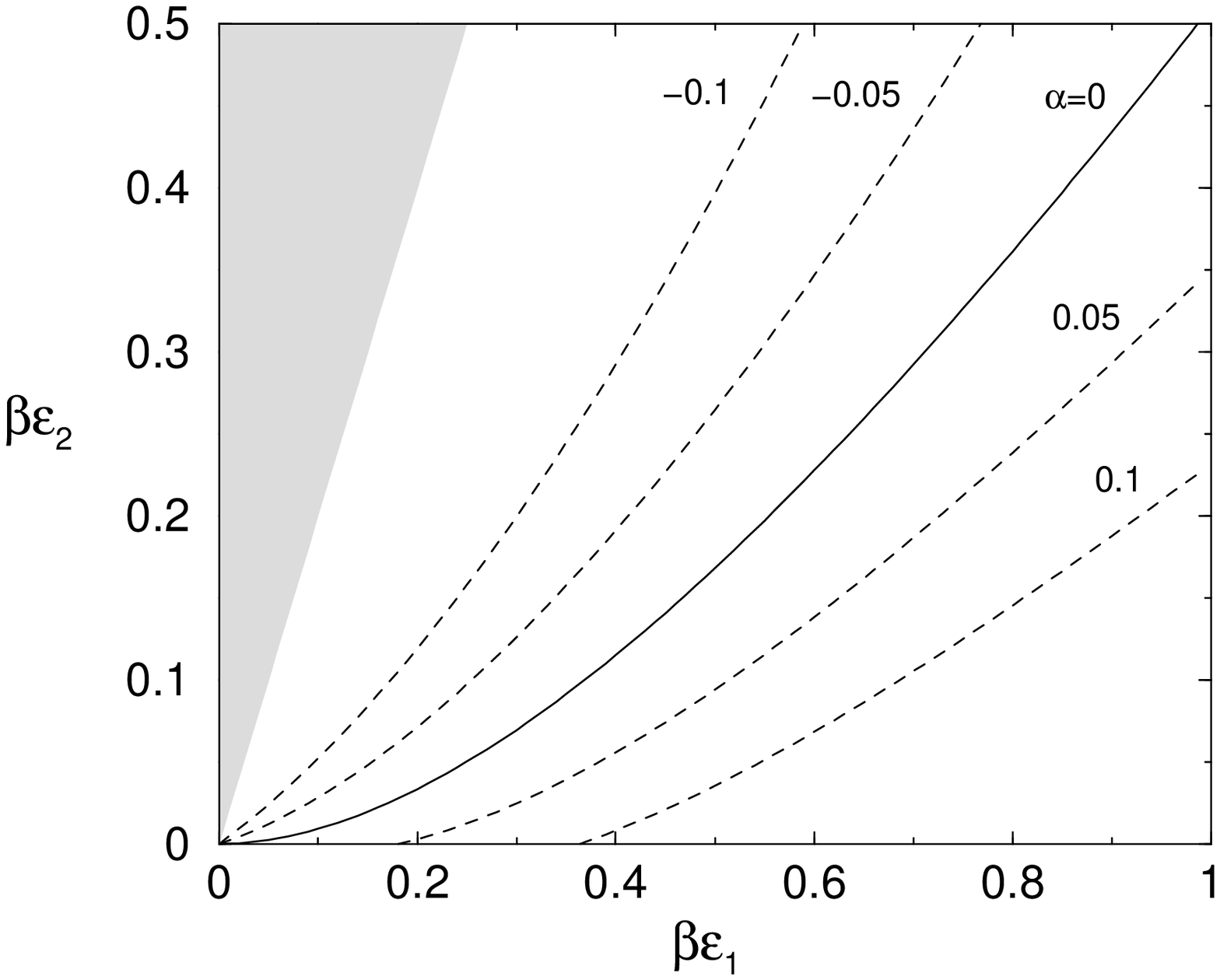}
\caption{\label{origalpha}  K. Shundyak, R. van Roij and P. van der Schoot}
\end{figure}

It is a straightforward exercise to calculate the second virial
coefficients $B_{\sigma\sigma'}$ averaged over all angles, from
the pair interactions $\phi_{\sigma\sigma'}(r)$ given
above\cite{O,VL}. In the Onsager limit $L\gg\Delta_2\geq\Delta_1$,
where terms of order $L\Delta^2$ can be ignored, one finds
\begin{eqnarray}
B_{11}&=&(\pi/4)L^2\Delta_1,\label{b11}\\
B_{12}&=&(\pi/4)L^2\left(\Delta_1+\frac{\Delta_2-\Delta_1}{2}\left(1-e^{-\beta
\epsilon_1}\right)\right),\label{b12}\\
B_{22}&=&(\pi/4)L^2\left(\Delta_1+\frac{\Delta_2-\Delta_1}{2}\left(2-e^{-2\beta
\epsilon_1}-e^{-\beta
\epsilon_2}\right)\right).\nonumber\\\label{b22}
\end{eqnarray}
These expressions can be used to map the model mixture of bare and
PEG-coated rods onto a mixture of hard rods with effective
hard-core diameters $D_1$ and $D_2$. We choose $D_1$ and $D_2$ to
be such that the like-like second virial coefficients of the
effective hard-core system are identical to $B_{11}$ and $B_{22}$
given in Eq. (\ref{b11}) and (\ref{b22}), respectively, i.e., we
impose that $B_{\sigma\sigma}=(\pi/4)L^2D_{\sigma}$. This yields
\begin{eqnarray}
D_1&=&\Delta_1, \nonumber\\
D_2&=&\Delta_1 + \frac{\Delta_2 -
\Delta_1}{2}(2-\exp(-2\beta\epsilon_1)-\exp(-\beta\epsilon_2)).\nonumber\\\label{d2}
\end{eqnarray}
One may verify that $D_2=\Delta_2$ in the limit that
$\beta\epsilon_i\rightarrow\infty$, as expected. We now also
impose that the cross virial coefficient of the effective
hard-core system equals $B_{12}$ given in Eq. (\ref{b12}). For
arbitrary $\epsilon_1$ and $\epsilon_2$ this requires a
nonadditivity parameter $\alpha$ such that
$B_{12}=(\pi/8)L^2(D_1+D_2)(1+\alpha)$, which yields
\begin{eqnarray}
\alpha=\frac{1}{d+1}\left[
2+\frac{2(d-1)(1-\exp(-\beta\epsilon_1))}{2-\exp(-2\beta\epsilon_1)
-\exp(-\beta\epsilon_2)} \right]-1,
\end{eqnarray}
with $d=D_2/D_1$ the effective diameter ratio of the rods. In Fig.
\ref{origalpha} we show the contour plot of the nonadditivity
parameter $\alpha$ as a function of the energy scales $\epsilon_1$
and $\epsilon_2$ for the effective diameter ratio $d=3.5$, which
is the value that we will use in our calculations below; other
values for $d$ produce similar contour plots. The grey area in
Fig. \ref{origalpha} is the regime deemed unphysical, with
$2\epsilon_1<\epsilon_2$. As one can see, for physically
reasonable values of $\epsilon_1$ and $\epsilon_2$ of the order of
$k_BT$ both positive and negative values for $\alpha$ are
possible, even when $\epsilon_1 > \epsilon_2$. The crossover from
positive to negative nonadditivity takes place, independently from
the value of $d$, when $\exp(-\beta\epsilon_2) =
2\exp(-\beta\epsilon_1) - \exp(-2\beta\epsilon_1)$, i.e., when
$\beta\epsilon_2\sim (\beta\epsilon_1 )^2$ if $\beta\epsilon_1 <
1$ and $\beta\epsilon_2 \sim \beta\epsilon_1 - \ln 2$ if
$\beta\epsilon_1 > 1$. Presuming that both $\beta \epsilon_1$ and
$\beta \epsilon_2$ are indeed of the order unity, we expect
$|\alpha|$ to be in the range $10^{-2} - 10^{-1}$. Such small
deviations from additivity are sufficient to qualitatively alter
the phase behavior of the rods, as we shall see next. In our
study, we from now on treat $\alpha$, $D_1$ and $D_2$ as
independent parameters. We investigate both the bulk and the
interfacial behavior of the effectively purely hard-core system,
in which the soft interactions are incorporated through the degree
of non-additivity $\alpha$.

\section{Density functional and method}
\label{nafunctsect} Consider a fluid of hard cylinders of two
different species $\sigma=1,2$ of diameter $D_{\sigma}$ and equal
length $L$ ($D_{\sigma}/L \rightarrow 0$) in a macroscopic volume
$V$ at temperature $T$ and chemical potentials $\mu_{\sigma}$. Let
${\bf r}$ denote the center-of-mass coordinate of a rod and ${\bf
\hat{\omega}}$ the orientation of the long axis. The interactions
between the $\sigma \sigma'$-pair of rods with coordinates
$q=\{{\bf r},{\bf \hat{\omega}}\}$ and $q'=\{{\bf r}',{\bf
\hat{\omega}}'\}$ are characterized by a hard-core potential,
which is the simple contact potential for rods of the same species
($\sigma=\sigma'$), whereas for unlike rods ($\sigma \neq
\sigma'$) it corresponds to interactions between hard rods of
diameter $(1+\alpha) D_1$ and $(1+\alpha) D_2$.

Within the second virial approximation and in the absence of
external potentials, the grand potential functional
$\Omega[\{\rho_{\sigma}\}]$ of the one-particle distribution
functions $\rho_{\sigma}({\bf r},{\bf \hat{\omega}})$ can be
written \cite{O,VL,SRPRE03} as
\begin{eqnarray}
\label{napot} \beta\Omega[\{\rho_{\sigma}\}]&=&\sum_{\sigma}\int
dq \rho_{\sigma}(q) \Big(\ln[\rho_{\sigma}(q)
L^2D_{\sigma}]-1-\beta\mu_{\sigma}\Big)\nonumber\\
&&-\frac{1}{2}\sum_{\sigma{\sigma}'} \int dq dq'
f_{\sigma{\sigma}'}(q;q') \rho_{\sigma}(q)\rho_{{\sigma}'}(q'),
\end{eqnarray}
with $\beta=(k_BT)^{-1}$ the inverse temperature, and
$f_{\sigma{\sigma}'} (q,q')$ the Mayer function, which equals $-1$
if the rods overlap and vanishes otherwise. Since we consider the
limit $D_{\sigma}/L \rightarrow 0$ for any $\sigma$, the relative
shape disparity of rods is characterized by the ratio $d=D_2/D_1$
of the diameters and the value of the nonadditivity $\alpha$.

The minimum conditions $\delta\Omega[\{\rho_{\sigma}\}]/
\delta\rho_{\sigma}(q)=0$ on the functional lead to the set of
nonlinear integral equations
\begin{eqnarray}
\label{nanonlinset} \ln[\rho_{\sigma}(q)L_{\sigma}^2D_{\sigma}]
-\sum_{{\sigma}'} \int d q' f_{\sigma{\sigma}'}(q;q')
\rho_{{\sigma}'}(q') =\beta\mu_{\sigma}
\end{eqnarray}
to be solved for the equilibrium distributions $\rho_{\sigma}(q)$.
These equations are identical to the Euler-Lagrange equations for
additive rods mixtures, and we can directly apply the method
developed earlier \cite{RMDP98,SRPRE03}. The structure of the bulk
phase diagram depends now on the value of the nonadditivity
parameter $\alpha$, and has to be determined first.

Since the bulk distribution functions of the isotropic and nematic
phase are translationally invariant, i.e., $\rho_{\sigma}({\bf
r},\hat{\omega})=\rho_{\sigma}(\hat{\omega})$, we can reduce Eq.
(\ref{nanonlinset}) to
\begin{equation}
\label{stathomnonadd}
\ln[\rho_{\sigma}(\hat{\omega})L_{\sigma}^2D_{\sigma}]+\sum_{\sigma'}
\int d\hat{\omega}'E_{\sigma{\sigma}'}(\hat{\omega},\hat{\omega}')
\rho_{{\sigma}'}(\hat{\omega}') =\beta\mu_{\sigma},
\end{equation}
with $E_{\sigma{\sigma}'}$ the excluded volume of a pair of
cylinders of species $\sigma$ and $\sigma'$ given by
\begin{eqnarray}
\label{exclvolnonadd}
E_{\sigma{\sigma}'}(\hat{\omega},\hat{\omega}')&=& -\int d{\bf
r}'f_{\sigma{\sigma}'}({\bf r},\hat{\omega};{\bf
r}',\hat{\omega}')\nonumber\\
&=&L^2(D_{\sigma}+D_{\sigma'})(1+\alpha (1 -
\delta_{\sigma,\sigma'})) |\sin\varphi|\nonumber\\
\end{eqnarray}
in terms of the angle $\varphi$ between $\hat{\omega}$ and
$\hat{\omega}'$, i.e., $\varphi=\arccos (\hat{\omega} \cdot
\hat{\omega}')$. Note that additional $O(LD^2)$ terms are being
ignored in Eq. (\ref{exclvolnonadd}), in line with the needle
limit ($D_{\sigma}/L \rightarrow 0$) of interest here. Given the
linear dependence of the excluded volume on $D_\sigma$, one can
see that
\begin{eqnarray}
E_{12}(\hat{\omega},\hat{\omega}')=
\frac{1}{2}(E_{11}(\hat{\omega},\hat{\omega}')
+E_{22}(\hat{\omega},\hat{\omega}'))(1+\alpha).
\end{eqnarray}
In some sense, $\alpha$ plays a similar role in the present
context as the so-called $\chi$ -parameter in the Flory theory of
polymer solutions on a lattice, where demixing is driven by direct
unfavorable nearest neighbor interaction between unlike species as
compared to that between like species.

Details of the numerical schemes to solve Eq.
(\ref{stathomnonadd}) have been discussed elsewhere
\cite{RMDP98,SRPRE03}. Here we use a nonequidistant $\theta$-grid
of $N_{\theta}=30$ points $\theta_i\in[0,\pi/2]$, where $1\leq
i\leq N_{\theta}$, in order to find the bulk distributions
$\rho_{\sigma}(\theta_i)$. Coexistence of different phases
$\{I,N_1,N_2\}$ is determined by imposing conditions of mechanical
and chemical equilibrium.

\section{Bulk phase diagrams}\label{nabulksect}
In Fig. \ref{nathinthickphase} we show both pressure-composition
(a) and density-density (b) representations of bulk phase diagrams
of thin-thick binary mixtures ($L_{\sigma}=L,D_2>D_1$) for the
diameter ratio $d=3.5$ at various values of the nonadditivity
parameter $\alpha$. In Fig. \ref{nathinthickphase}(a) the
composition variable $x=n_2/(n_1+n_2)$ denotes the mole fraction
of thick rods, $n_{\sigma}=\int d\hat{\omega}
\rho_{\sigma}(\hat{\omega})$ is the number density of species
$\sigma$, and $p^{\ast}=(\pi/4)\beta pL^2D_1$ is a dimensionless
bulk pressure. Note that the $IN$ coexistence pressure $p_{thin}$
and $p_{thick}$ of the pure thin ($x=0$) and pure thick ($x=1$)
system are given by $(\pi /4) \beta p_{thin}L^2D_1= (\pi /4) \beta
p_{thick}L^2D_2=14.045$, i.e., $p_{thick}=p_{thin}/d$, and that
the tie-lines connecting coexisting phases are horizontal in the
$p-x$ representation of Fig. \ref{nathinthickphase}(a). This
representation is convenient for theoretical analysis, whereas the
densities (volume fractions) of thin and thick rods are
experimental control parameters \cite{F}. For this reason the same
phase diagrams of thin-thick binary mixtures are shown in Fig.
\ref{nathinthickphase}(b) in the density-density representation,
with $n_1^{*}=n_1L^2D_1(\pi/4)$ and $n_2^{*}=n_1L^2D_2(\pi/4)$
being the dimensionless bulk number densities of thin and thick
rods, respectively. In this representation the tie-lines,
indicated by the dotted lines, are no longer horizontal.

The structures of the bulk phase diagrams for various $\alpha$
show a striking similarity with the bulk phase diagrams of
additive binary mixtures of thin and thick hard rods
\cite{SRPRE03}. At low pressures (or low densities) the phase
diagrams show an isotropic ($I$) phase, and at higher pressures
(or densities) one ($\alpha<0.07$) or two ($\alpha \geq 0.07$)
nematic phases ($N_1$ and $N_2$). For $\alpha<0.07$ the phase
diagram is spindle-like, and the only feature is a strong
fractionation at coexistence, such that the nematic phase is
relatively rich in thick rods and the isotropic phase in thin
ones. Although the nonadditivity modifies the fractionation gap,
the reason behind it remains the same: the relatively large
excluded volume in interactions of the thick rods makes them more
susceptible to orientational ordering
\cite{LCHDJCP84,RMJCP96,RMDP98}. As a general tendency, the
fractionation at isotropic-nematic coexistence becomes stronger
for increasing values of $\alpha$.

For $\alpha>0.06$ the bulk phase diagram develops nematic-nematic
($N_1N_2$) coexistence in a pressure regime $p>p_t$, with $p_t$
the triple-point pressure.  Using the simple Gaussian ansatz for
one-particle distribution functions, one can demonstrate that the
packing entropy does not play a role in nematic demixing in our
system, similar to the case of additive mixtures \cite{VLJPC93}.
Although it is known that the functional form of
$\rho_{\sigma}(\hat{\omega})$ is not Gaussian even at high
densities, an analysis of the exact high-density distribution
functions confirmed such a mechanism of nematic demixing
\cite{RMDP98}. On this basis we assume it to be valid at arbitrary
high pressure in our system, and expect the structure of the bulk
phase diagrams to be similar to those of additive mixtures. In
particular, for $\alpha=0.07$ nematic remixing is observed at a
sufficiently high pressure, as illustrated in Fig.
\ref{nathinthickphase}. The consolute point, at which the density
and composition difference between the coexisting nematic phases
vanishes, is indicated by ($\ast$). For $\alpha=0.1$, limitations
of the numerical scheme \cite{SRPRE04} do not allow us to
determine whether or not remixing takes place at high enough
pressures. We note that in the limit of very high pressures, where
the rods increasingly align themselves, both end corrections and
higher order virials need to be taken into account for an accurate
description of the phase behavior. On the other hand, in analogy
with additive mixtures, one expects that critical values of
$\alpha$ and $d$, beyond which the nematic demixing does not take place at
arbitrary high densities, exist \cite{RMDP98}.

\begin{figure}[!t]\centering
\includegraphics[width=8.5cm]{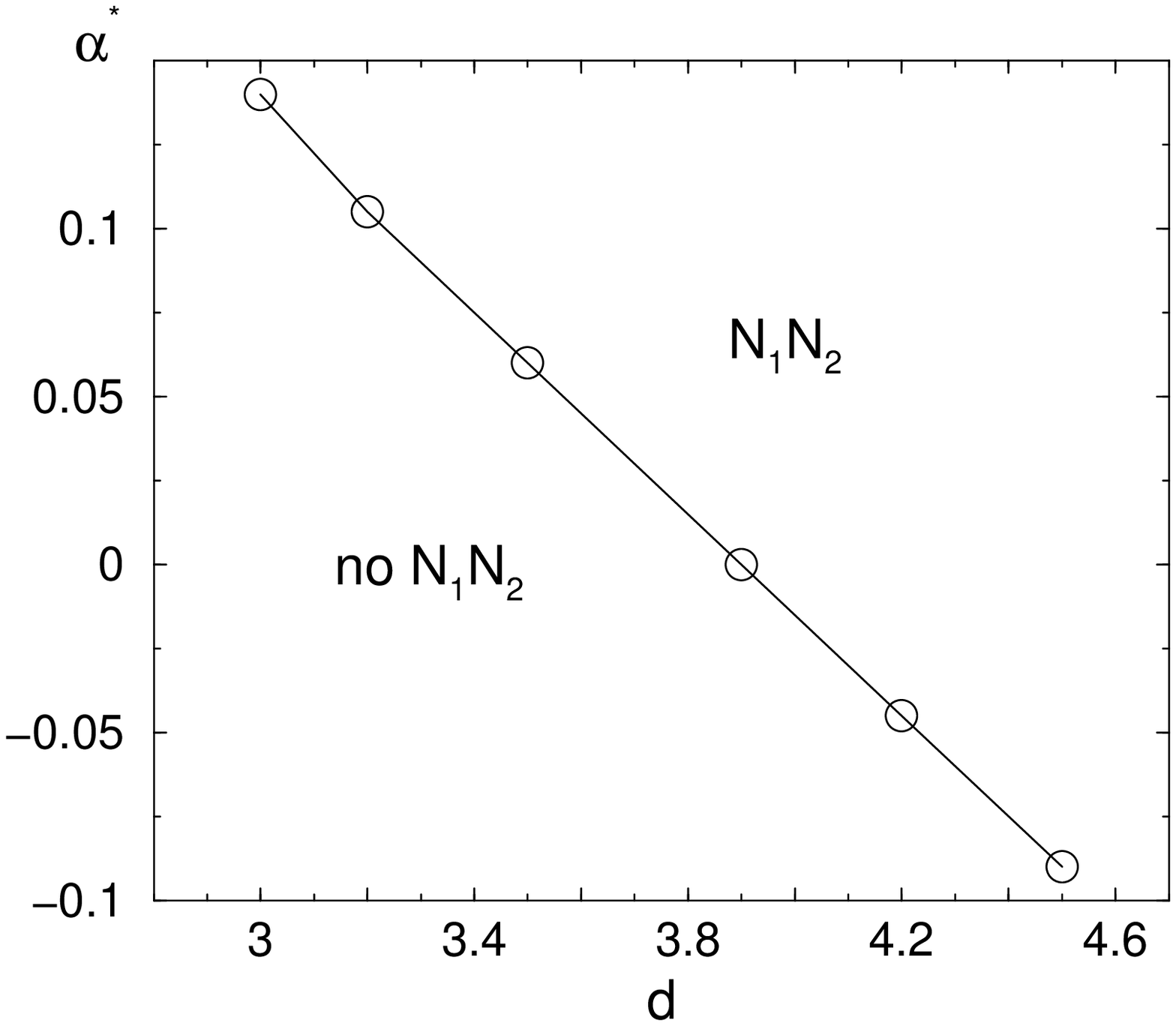}
\caption{\label{naalpha}  K. Shundyak, R. van Roij and P. van der Schoot }
\end{figure}

In order to characterize the amount of nonadditivity in the
excluded volume interactions which leads to significant structural
modification of the phase diagram (i.e. nematic demixing), we
explore various thin-thick mixtures of different values of
$\alpha$ and $d$, and determine the value of $\alpha^{\ast}$ for
which the pressure of the nematic-nematic consolute point and the
triple point pressure coincide. Results of our studies are
presented in Fig. \ref{naalpha}. For $\alpha< \alpha^{\ast}$ (at
fixed $d$) the $N_1N_2$ phase separation is not detected, and for
$\alpha \geq \alpha^{\ast}$ there is $N_1N_2$ coexistence in the
phase diagram. It is evident that in the interval $d \in [3.5,
4.2]$ even a small nonadditivity $|\alpha|<5-7 \%$ may induce or
suppress the $N_1N_2$ demixing transition.

One might surmise that the linearity of the function
$\alpha^{\ast}(d)$ within the explored range of $\alpha$ reflects
the linearity of the excluded volume
$E_{12}(\hat{\omega},\hat{\omega}')$ in terms of $d$ and $\alpha$,
because it drives the nematic-nematic phase separation. The
mapping of the non-additive to the additive case is not trivial,however,
since the density distributions $\rho_\sigma(\hat{\omega})$ 
depend on $\alpha$ implicitely.
Nonetheless, direct comparison of the bulk phase diagrams of the nonadditive mixture
with $d=3.5$ and $\alpha=0.07$ and the additive mixture with, for instance, 
$d=4.0$\cite{SRPRE03} shows close values of the
fractionation gap at the $N_1N_2$ coexistence. Further evidence
for similarity of these systems in the high density regime will be
demonstrated in our analysis of their interfacial properties
presented next.

\section{Interfaces}\label{nainterfsect}
Free planar interfaces between various coexisting bulk phases can
be studied similar to the interfaces of additive mixtures
\cite{SRJPCM2001,SRPRL02,SRPRE03}. We focus on the nonadditive
thin-thick mixture characterized by $d=3.5$ and $\alpha=0.07$. The
nematic director $\hat{n}$ of the asymptotic nematic bulk phase(s)
can, in general, have a nontrivial tilt angle
$\theta_{t}=\arccos(\hat{n} \cdot\hat{z})$ with respect to the
interface normal $\hat{z}$. In the present calculations we restrict
attention to $\theta_{t}=\pi/2$, i.e., $\hat{n} \perp \hat{z}$. As
we have checked, this geometry is thermodynamically favorable
because of its minimal surface tension.

Similar to the studies of additive mixtures, we use the planar
symmetry of the interfaces and assume the distribution functions
to be uniaxially symmetric with respect to the director, i.e.
$\rho_{\sigma}({\bf r}, \hat{\omega}) =\rho_{\sigma}(z,\theta)$,
which reduce Eqs. (\ref{nanonlinset}) to
\begin{eqnarray}
\label{narho0} \beta\mu_{\sigma}&=&\ln[\rho_{\sigma}(z,\theta)
L_{\sigma}^2D_{\sigma}]+\sum_{{\sigma}'}\int dz' d\theta'
\sin\theta' \nonumber\\
&&\times \mathcal{K}_{\sigma{\sigma}'} (z-z',\theta,\theta')
\rho_{\sigma'}(z',\theta'),
\end{eqnarray}
with $\mathcal{K}_{\sigma{\sigma}'}
(z-z',\theta,\theta')=-\frac{1}{2\pi} \int d\varphi d\varphi'
dx'dy' f_{\sigma{\sigma}'}(q,q')$. We solve Eq. (\ref{narho0}) in
order to determine uniaxially symmetric nonuniform distributions
$\rho_{\sigma}(z,\theta_i)$ using an equidistant $z$-grid of
$N_z=200$ points in the interval $z\in [-5L,5L]$, and
corresponding bulk distributions $\rho_{\sigma}(\theta_i)$ as
boundary conditions. Further details of the numerical calculations
were discussed elsewhere \cite{SRPRE03}.

The $IN_1$ and $N_1N_2$ interfaces are found to be smooth and
monotonic, in the sense that the profiles of the nematic uniaxial
order parameters $S_{\sigma}(z)$ and the densities $n_{\sigma}(z)$
change monotonically from the bulk values in the $I$ ($N_1$) phase
to those in the $N_1$ ($N_2$) phase. The correlation length
$\xi_{N_1}$ of the bulk $N_1$ phase at the triple-phase
coexistence (as well as $\xi_{I}$ and $\xi_{N_2}$ for the $I$
phase and the $N_2$ phase, respectively) can be extracted from the
asymptotic decay of the one-particle distributions $\rho_\sigma
(z,\theta)$ to their bulk values
$\rho_{\sigma}^{N_1}(\hat{\omega})$, since the deviation $\delta
\rho_{\sigma}(z,\hat{\omega})=\rho_{\sigma}(z,\hat{\omega}) -
\rho_{\sigma}^{N_1}(\hat{\omega})$ is of the form \cite{SRPRE03}
\begin{eqnarray}
\label{asymptdens}  \delta \rho_{\sigma}(z,\hat{\omega})=
A_{\sigma} (\hat{\omega}) \exp (-z/ \xi_{N_1}),\;\; z\rightarrow
\infty.
\end{eqnarray}
Interestingly, we find that $\xi_{N_1}/L=0.49 \pm 0.02$ is the
same as for the additive mixture with $d=4.0$ \cite{SRPRE03},
which has a virtually identical phase diagram.

\begin{figure}[!t]\centering
\includegraphics[width=8.5cm]{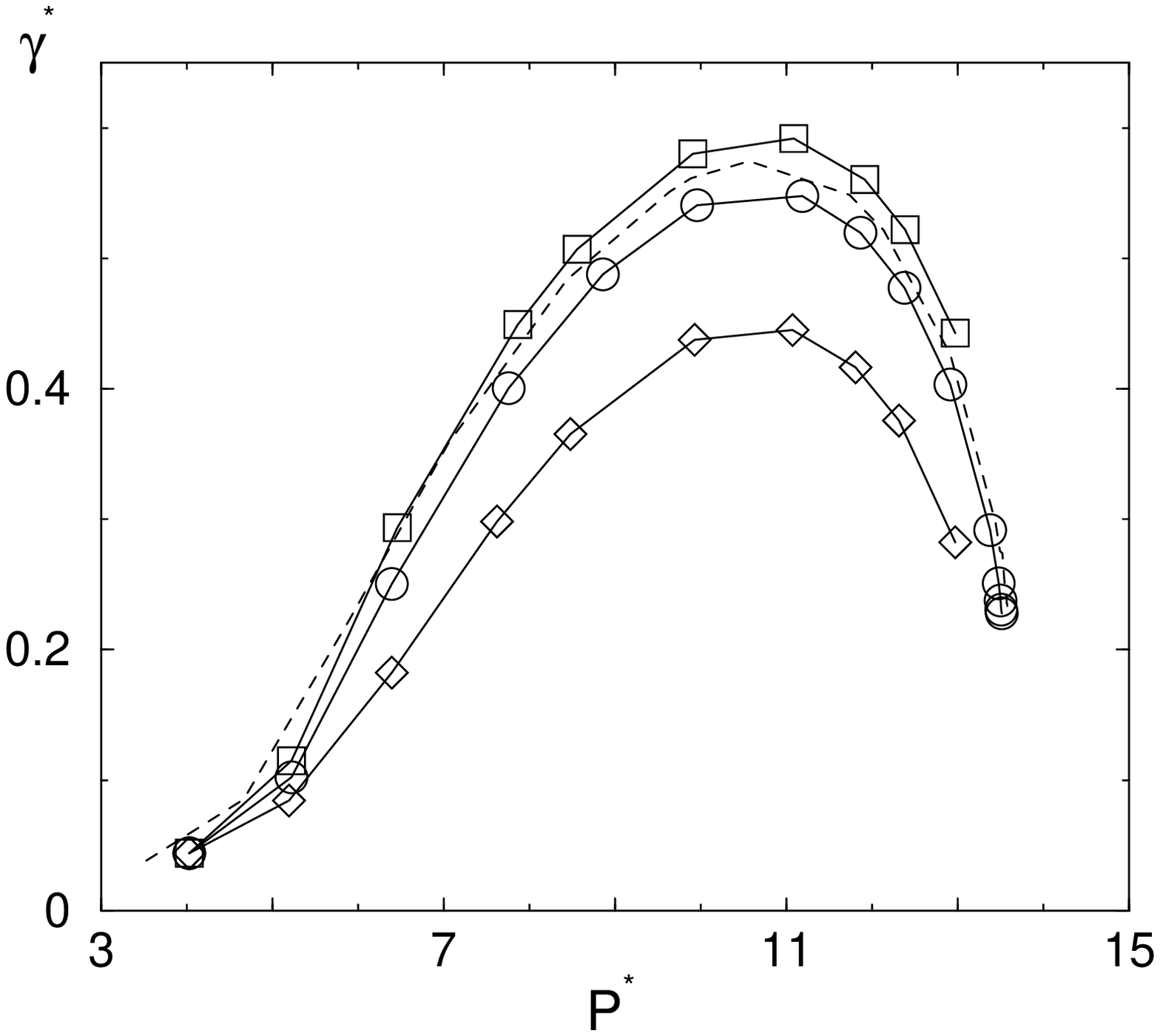}
\caption{\label{natensionall}  K. Shundyak,R. van Roij and P. van der Schoot}
\end{figure}

The properties of the $IN_2$ interfaces depend strongly on the
pressure difference with the triple point ($IN_1N_2$ phase
coexistence). As it is demonstrated in Fig. \ref{natensionall},
the surface tension of the $IN_2$ interface shows a non-monotonic
dependence on the bulk pressure $p$, and strongly correlated with
the fractionation at the $IN_2$ coexistence. Upon increasing the
nonadditivity, the surface tension $\gamma_{IN_2}^{*}(p)$ grows,
again indicating that $\alpha$ plays a role similar to the
diameter ratio $d$. For comparison we have included
$\gamma_{IN_2}^{*}(p)$ for an additive thin-thick mixture with
$d=4.0$, which is again quite close to the results for the
nonadditive mixtures with $d=3.5$ and $\alpha=0.07$ and $0.1$.

The microscopic thickness $t$ of the interface is defined as
$t=|z_+-z_-|$ where $z_\pm$ are solutions of ${n_1}'''(z)=0$, and
a prime denotes differentiation with respect to $z$. As this
equation has a set of solutions in every interfacial region, we
choose for $z_\pm$ the outermost ones, i.e., the ones nearest to
the bulk phases. The density of thin rods is a convenient
representation of structural changes within the interface, since
they have a smaller excluded volume and a non-vanishing
concentration in both coexisting phases. This criterion provides a
single measure for the thickness of both monotonic and
non-monotonic profiles, with and without a thick film in between
the asymptotic bulk phases at $z\rightarrow \pm \infty$. The
interfacial width for the one-component $IN$ interface is, with
the present definition, given by $t/L=0.697$.

\begin{figure}[!t]\centering
\includegraphics[width=8.5cm]{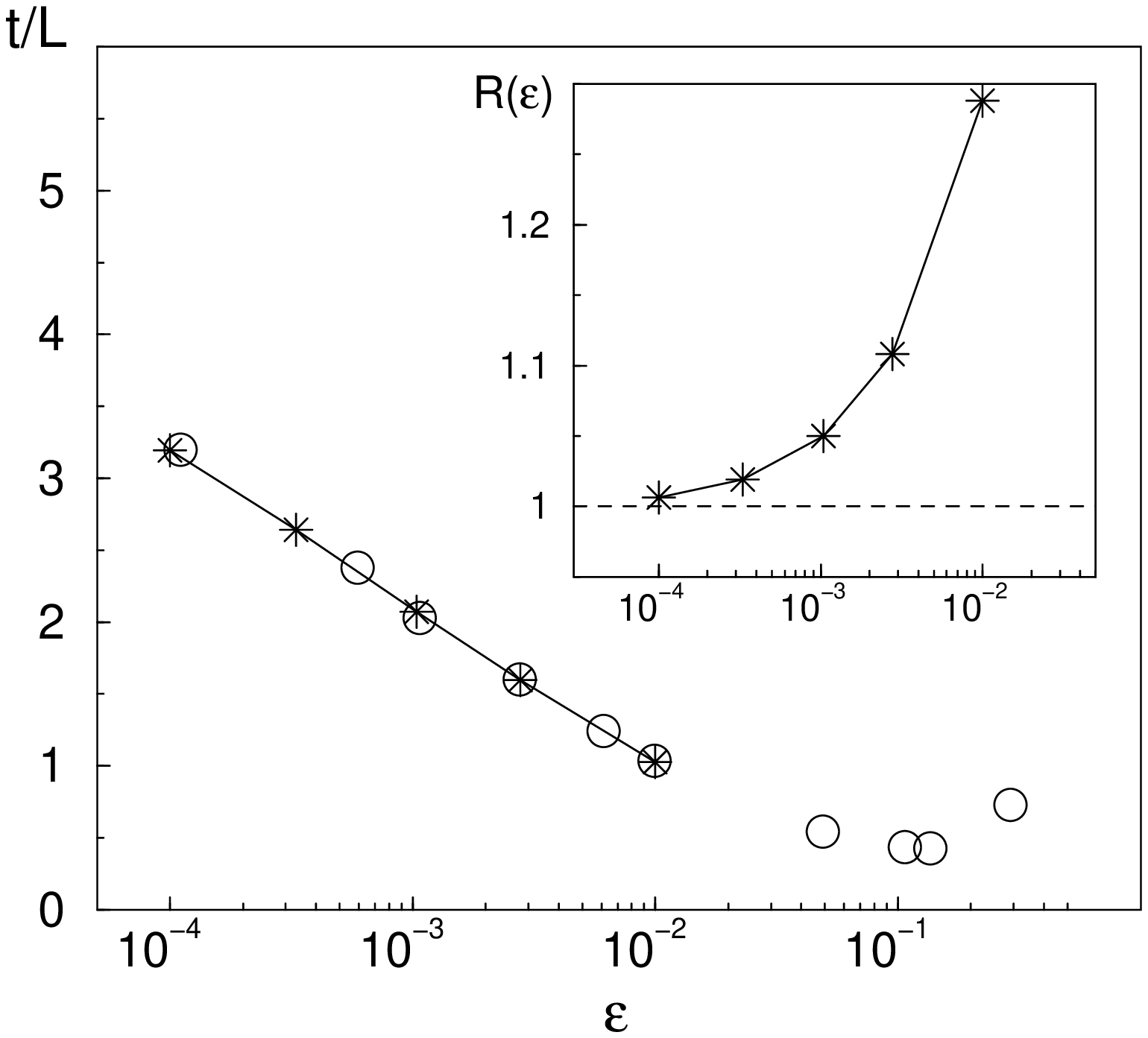}
\caption{\label{nafilms}  K. Shundyak, R. van Roij and P. van der Schoot}
\end{figure}

\begin{figure}[!t]\centering
\includegraphics[width=8.5cm]{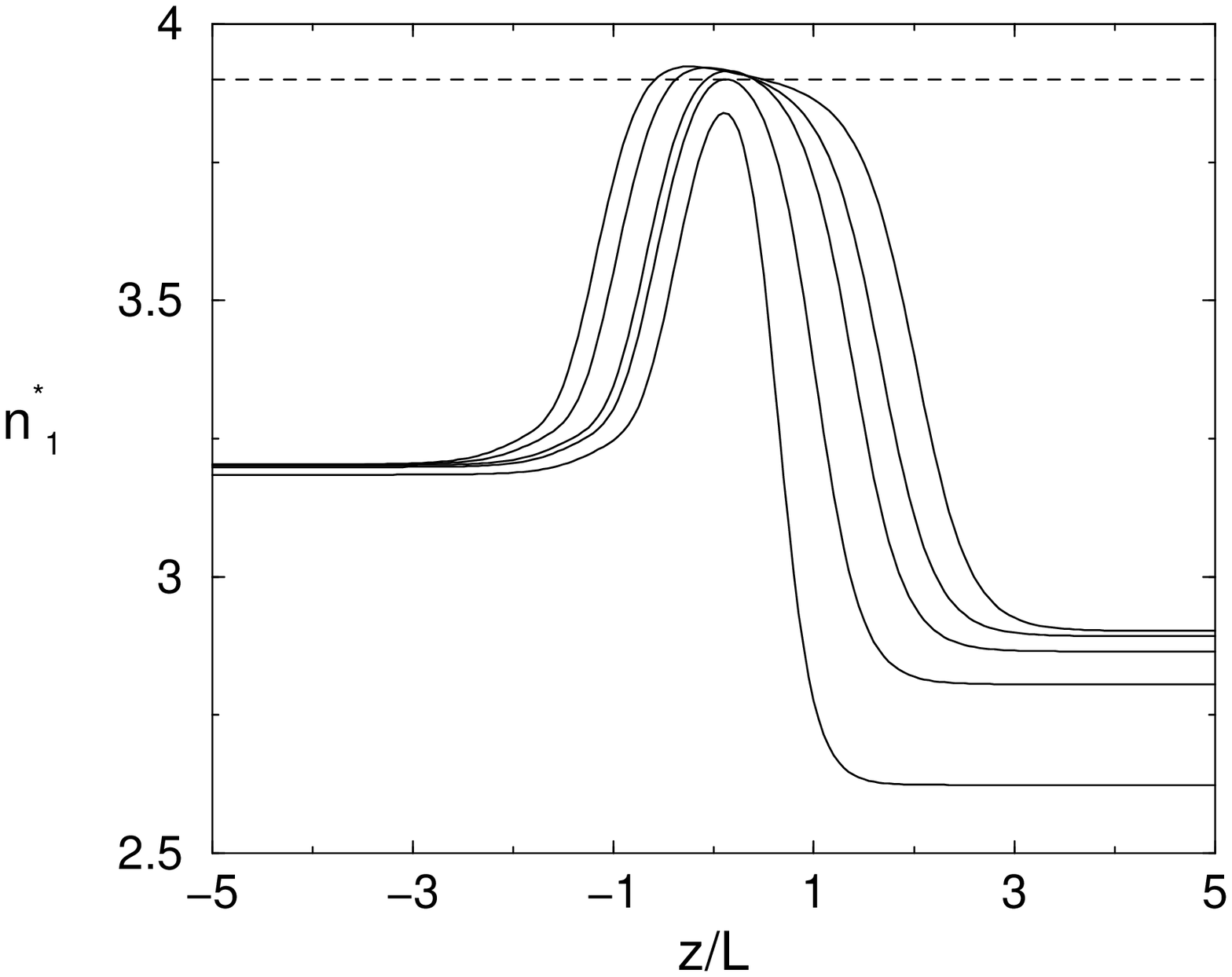}
\caption{\label{nan1}  K. Shundyak, R. van Roij and P. van der Schoot}
\end{figure}

The thickness of the $IN_2$ interface was found to diverge upon
approach of the triple-point pressure $p_t$. This can be seen in
Fig. \ref{nafilms}, where $t/L$ is plotted as a function of the
dimensionless undersaturation $\epsilon=1-p/p_t$, which is a
convenient measure of the pressure difference with the triple
point. The nature of the film can be analyzed from the density
profiles $n_1(z)$ of the $IN_2$ interface (or equivalently
$S_{\sigma}(z)$, or $n_2(z)$). In Fig. \ref{nan1} the profiles of
$n_1(z)$ are shown at several values of the undersaturation
$\epsilon$. The asymptotic densities at $z \rightarrow \pm \infty$
in Fig. \ref{nan1} are those of the coexisting $I$ and $N_2$ bulk
phases (at the corresponding $\epsilon$). For $\epsilon
\rightarrow 0$ the value of $n_1(z)$ in the film approaches the density
of thin rods of the bulk triple point $N_1$ phase, indicated by
the dashed line in Fig. \ref{nan1}. However, the undersaturation
$\epsilon=10^{-4}$ is yet too large to be in the asymptotic
thick-film regime. The same identification can be made for
$n_2(z)$ and $S_{\sigma}(z)$, and on this basis we conclude that
the wetting phenomenon under consideration is complete triple
point wetting of the free $IN_2$ interface by the $N_1$ phase. The
similarity with complete wetting of the $IN_2$ interface by the
$N_1$ phase in additive thin-thick mixture with $d=4.0$ is again
rather striking, as is clear from Fig. \ref{nafilms}, where the
thickness of the $IN_2$ interface for additive rods is indicated
by ($\circ$).

Since one expects, for the short-range interactions of interest
here, that the thickness of the wetting $N_1$ film in the $IN_2$
interface diverges as $t \sim -\xi_{N_1} \ln \epsilon$ for
$\epsilon \rightarrow 0$ \cite{E79}, the value of the correlation
length of the bulk $N_1$ phase can be extracted,
$\xi_{N_1}=0.49\pm 0.02$, which is consistent with the value determined 
earlier from the decay of $\rho_\sigma(z,\theta)$ into the
bulk $N_1$ phase.

The analysis of the structural properties of the $IN_2$ interface
can be complemented by studies of the ratio of surface tensions
\begin{eqnarray}
\label{naR} R(\epsilon)=\frac{\gamma_{IN_2}(\epsilon
)}{\lim_{p\downarrow p_t}(\gamma_{IN_1}+\gamma_{N_1N_2})},
\end{eqnarray}
as presented in the Inset in Fig. \ref{nafilms}. It is clear that
upon approach of triple-point coexistence $\lim_{\epsilon
\rightarrow 0}R(\epsilon) =1$, which implies a vanishing contact
angle. This provides the thermodynamic proof of complete
triple-point wetting. The $\epsilon$-dependence of $R$ again
reveals that $\epsilon =10^{-4}$ is too large to be in the
asymptotic thick-film regime.

\section{Summary and discussion}\label{nasummarysect}
In this paper we have explored the bulk phase diagrams and the
interfacial properties of the nonadditive mixtures of thin and
thick hard rods. The nonadditivity was introduced in an attempt to
effectively capture some of the effects of soft interactions
between them, having in mind mixtures of bare and PEG-coated fd
virus particles in aqueous suspension \cite{PVGJFcondmat}. We
showed that the effective hard-core diameter of the unlike
interactions, $\frac{1}{2}(D_1+D_2)(1+\alpha)$ with $D_1$ and
$D_2$ the effective diameter of the like interactions, can easily
be smaller or larger than the additive case,
$\frac{1}{2}(D_1+D_2)$, by more than a few per cent.

\begin{figure}[!t]\centering
\includegraphics[width=8.5cm]{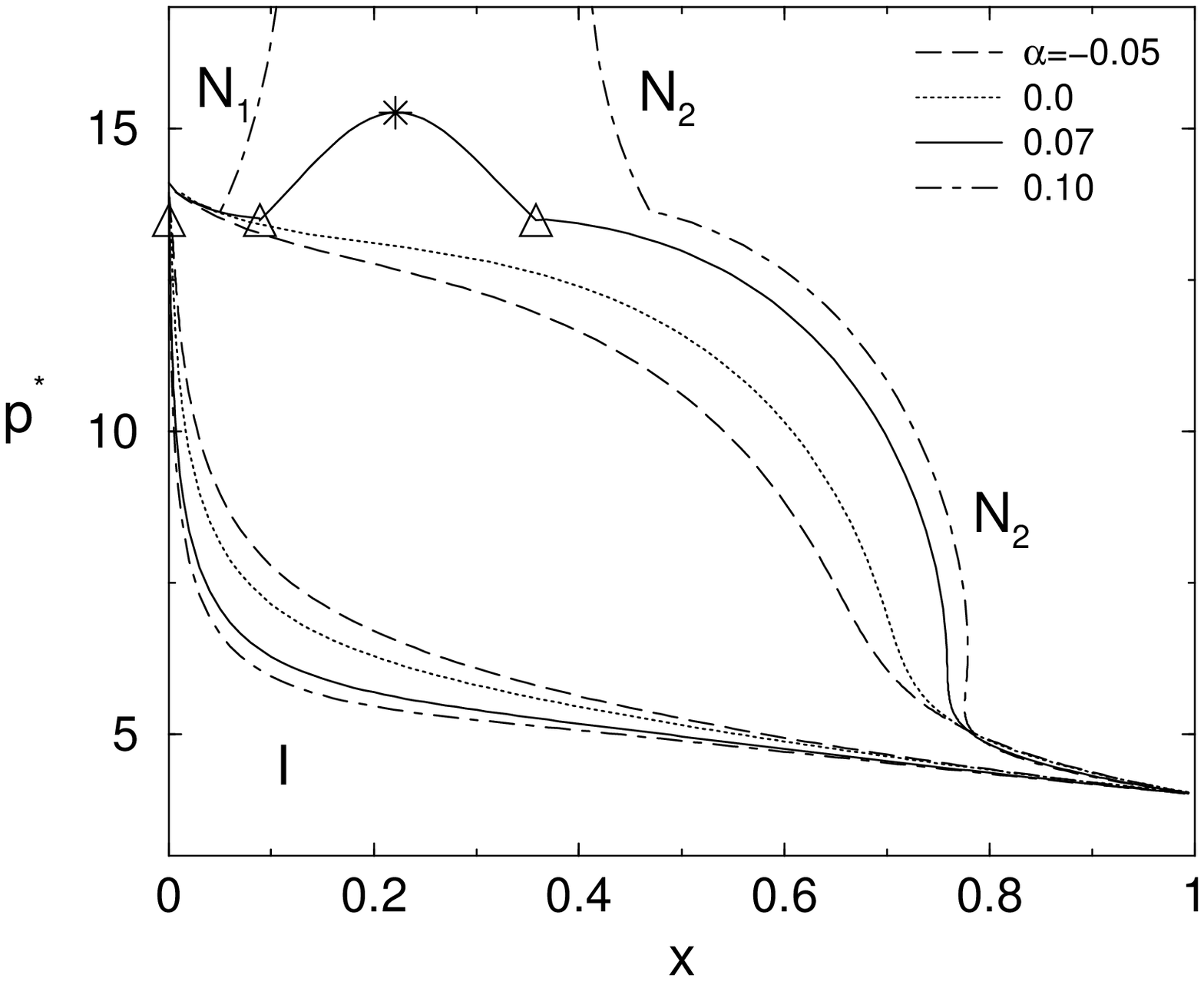}
\includegraphics[width=8.5cm]{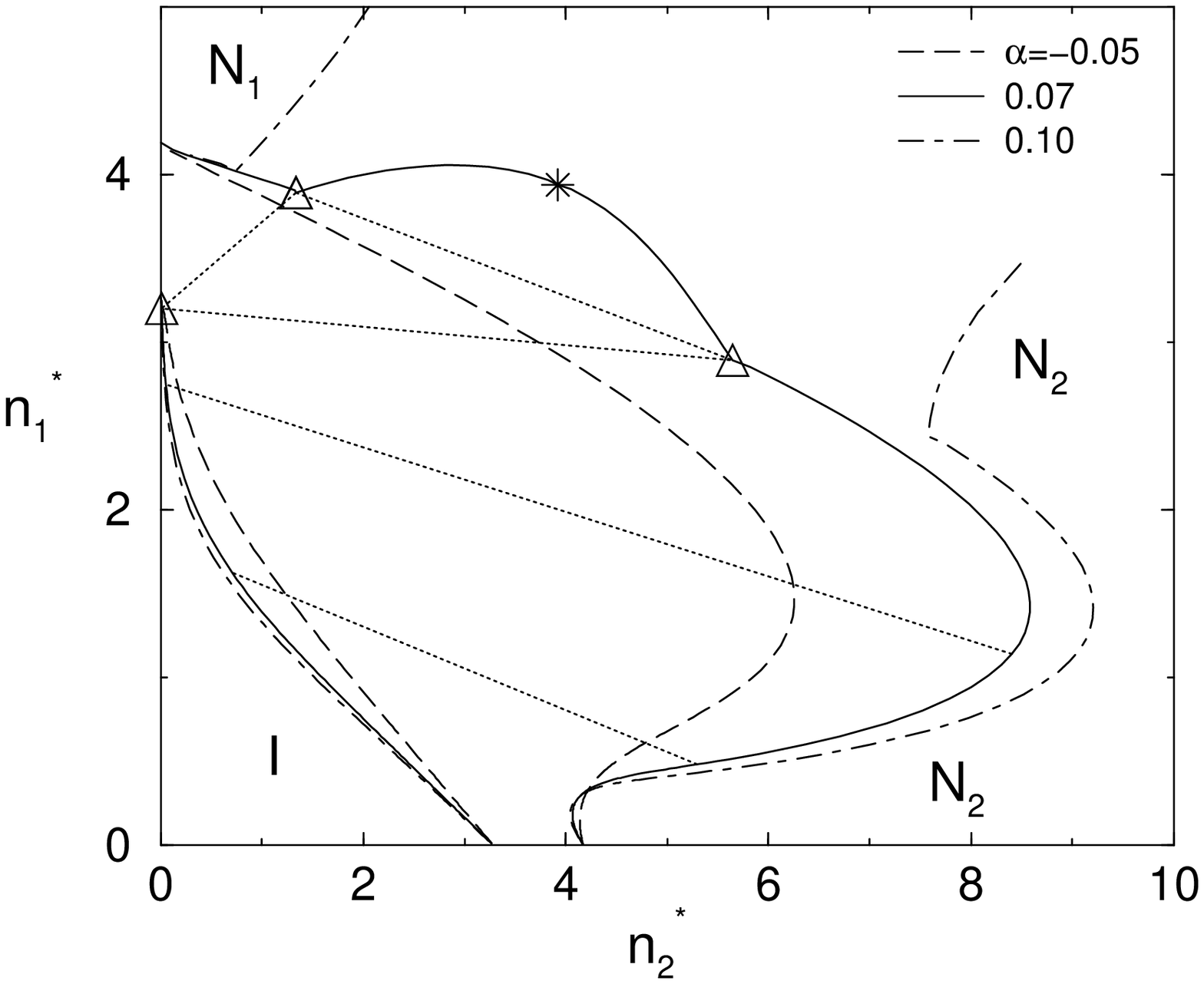}
\caption{\label{nathinthickphase}  K. Shundyak, R. van Roij and P. van der Schoot}
\end{figure}

As is illustrated in Fig. \ref{nathinthickphase}, a small amount
of nonadditivity $\alpha
>0$ can stabilize the high-density nematic-nematic phase coexistence, even if
it is only metastable for an additive mixture with the same
diameter ratio. However, the experimentally observed lower
critical point of the nematic-nematic demixing transition
\cite{PVGJFcondmat} could {\em not} be reproduced by incorporating
nonadditivity into the theory. We suggest, therefore, that further
theoretical studies of this system should consider in more detail
the impact in particular of a finite bending flexibility, beyond
the ground-state approximation \cite{semenov-subbotin}. Another
issue that needs to be resolved is the (elastic) response of the
polymer coat to volume exclusion between the rods, an aspect
completely ignored in our analysis.

We present results of bulk and interfacial 
calculations for the specific diameter ratio $d=3.5$, motivated by 
the experimental parameters \cite{PVGJFcondmat}. Other values of
$d$ lead to similar conclusions. We find the bulk phase diagrams of nonadditive binary mixtures to
show a large similarity with those of the additive mixtures of
larger diameter ratio. This is most likely related to the linear
dependence of the rod-rod excluded volume on both the diameter
ratio $d$ and the nonadditivity $\alpha$ although it is not
clear whether there is an exact mapping linking non-additive and
additive hard-rod mixtures. We also found that many if not all of
the interfacial phenomena that we studied are similar to those of
additive mixtures with a larger diameter ratio. Similar to the
interfaces between different bulk phases in additive mixtures, the
$IN_1$ and $N_1N_2$ interfaces are smooth and monotonic, whereas
the $IN_2$ interface exhibits complete wetting by the $N_1$ phase
upon approach of the triple phase coexistence. The complete
triple-point wetting scenario was confirmed by (i) the logarithmic
divergence of the thickness of the $N_1$ film with vanishing
undersaturation, and (ii) the surface tension ratio
$\lim_{\epsilon \rightarrow 0} R=1$. Such a similarity between
properties of additive and nonadditive mixtures may represent a
significant difficulty to distinguish these in experiments.

\begin{acknowledgments}
It is a pleasure to thank Kirstin Purdy, Henk Lekkerkerker, and
Seth Fraden for stimulating discussions, and Seth Fraden and
Kirstin Purdy for sharing unpublished experimental results with
us. This work is part of the research program of the 'Stichting
voor Fundamenteel Onderzoek der Materie (FOM)', which is
financially supported by the 'Nederlandse organisatie voor
Wetenschappelijk Onderzoek (NWO)'.
\end{acknowledgments}

\widetext
\newpage
FIGURE CAPTIONS
\begin{enumerate}
\item
Interaction potentials
$\phi_{11}(r)$, $\phi_{12}(r)$, $\phi_{22}(r)$ between bare-bare
(solid), bare-coated (dashed), coated-coated (dotted line) rods,
respectively, as a function of the (shortest) distance $r$ between
the axes of the two rods.

\item
Contour plot of the nonadditivity parameter $\alpha$ as a function of the
square-shoulder values $\epsilon_1$ and $\epsilon_2$ for the
effective diameter ratio $D_2/D_1=d=3.5$. The grey area denotes
the nonphysical region $\epsilon_2>2\epsilon_1$ (see text).

\item
(a) Bulk phase diagrams
of binary thin-thick mixtures (diameter ratio $d=3.5$) for
different nonadditivity parameter $\alpha$ in the $p-x$
representation, with $p^{\ast}=(\pi/4)\beta pL^2D_1$ the
dimensionless pressure, and $x$ the mole fraction of the thicker
rods. We distinguish the fully symmetric isotropic phase ($I$) and
orientationally ordered nematic phases ($N_1$ and $N_2$). For the
nonadditivity parameter $\alpha=0.07$ the $IN_1N_2$ triple phase
coexistence is marked by ($\triangle$), and the $N_1N_2$ critical
point by ($\ast$). (b) The same phase diagrams in density-density
representation, where $n_1^{*}=n_1L^2D_1(\pi/4)$ and
$n_2^{*}=n_2L^2D_2(\pi/4)$ are the dimensionless bulk number
densities of thin and thick rods, respectively. The tie-lines
connect coexisting state points.

\item
Nonadditivity parameter
$\alpha^{\ast}$ at which the consolute point and triple point
coincide for various values of the diameter ratio $d$. For
mixtures, characterized by $\alpha(d) < \alpha^{\ast}(d)$ the
$N_1N_2$ demixing is not detected.

\item
Dimensionless surface tension
$\gamma_{IN_2}^{*}=\beta \gamma_{IN_2} LD_1$ of $IN_2$ interfaces
as a function of dimensionless pressure $p^{*}=\beta p
L^2D_1(\pi/4)$ for thin-thick mixtures with diameter ratio $d=3.5$
and different values of nonadditivity $\alpha=0.0$ $(\diamond)$,
$0.07$ $(\circ)$ and $0.1 (\square)$. The dashed line indicates
$\gamma_{IN_2}^{*}(p^*)$ for the additive thin-thick mixture of
diameter ratio $d=4.0$.

\item
Thickness $t/L$ as a function of
the undersaturation $\epsilon=1-p/p_t$ from the triple point
pressure $p_t$ for thin-thick mixtures with diameter ratio $d=3.5$
and nonadditivity $\alpha=0.07$ $(\ast)$. For comparison we show
thickness of the $IN_2$ interface of the additive mixtures for
$d=4.0$ ($\circ$). The inset shows the surface tension ratio $R$
[see Eq. (\ref{naR})] as a function of the triple point
undersaturation $\epsilon$.

\item
Density profiles of the thin rods
$n_1^*(z)$ in the $IN_2$ interface for diameter ratio $d=3.5$ and
nonadditivity $\alpha=0.07$ at triple point undersaturations
$\epsilon=1-p/p_t=10^{-2}, 10^{-2.5}, 10^{-3}, 10^{-3.5},
10^{-4}$ (from bottom to the top curve). The bulk $I/N_2$ phase is at $z \rightarrow
-\infty/\infty$. The dashed line $n_1^* =3.977$ represents the
bulk density of thin rods in the triple point $N_1$ phase. These
profiles indicate the formation of a wetting $N_1$ film in the
$IN_2$ interface.

\end{enumerate}

\newpage

%\newpage

%\newpage

%\newpage

%\newpage

%\newpage

%\newpage

\end{document}